# Confirmation of some formulas related to spin coherence time

Yuri Orlov (Cornell University)

**Abstract**

This paper considers two sets of formulas related to a Spin Coherence Time (SCT) case with only vertical oscillations in a purely electric ring—the first derived by the author for the field index $m > 0$ and the second by Ivan Koop for the field index $m = 0$. I argue that a continuous transition can exist from one set to the other (contrary to appearances), and assume that a necessary condition for the transition is that both sets of formulas follow from the same equation. I demonstrate that they do follow when one takes into account that the first set of formulas holds only for times much larger than the period of the vertical oscillations. This demonstration confirms the correctness of the formulas and the equation.

**Introduction**

*A note on notations*. In what follows, the notations are unchanged from the original appearance of this paper in 2013 as a Storage Ring EDM Collaboration note [1]: TT refers to a 2012 workshop talk [2] and TN to a 2013 EDM note adding important details to that talk [3]; "Trento" broadly refers to either; and Ivan Koop's notation $\Delta x$ is used instead of my $\delta x$.

There exists an apparent gap between the formulas of Ivan Koop [4] and myself [2], [3] for $\Delta\gamma/\gamma$ and $\Delta x/R$ in an SCT case when only vertical oscillations are present.

More specifically, there appears to be no continuous transition from the formulas I have derived for the field index $m > 0$ to the formulas Koop has derived for the field index $m = 0$.

I believe that such a transition can in fact exist. The reason is that my formulas hold only for times much larger than the period of vertical oscillations, that is,

$$t \gg T_y = \frac{2\pi}{\omega_C \sqrt{m}}, \quad 1/N^2 \ll m \ll 1, \tag{1}$$

where $N$ is the number of revolutions. This restriction creates a finite interval, $\Delta m$, inside which the required continuous transition can occur.

One could formally demonstrate the transition by fixing some big non-infinite $t = t_{max}$ and calculating $\Delta\gamma = \Delta\gamma(m; t_{max})$, $\Delta x = \Delta x(m; t_{max})$ for different $m$'s but the same $t_{max}$. For any non-infinite $t_{max}$, one would then look for a continuous transition of $\Delta\gamma$, $\Delta x$ from my values to Koop's at $m \to 0$, inside the interval $\Delta m \sim (2\pi/t_{max}\omega_C)^2$, beginning from some $m$ that satisfies condition (1) and going down to $m = 0$. The bigger $t_{max}$ is, the sharper the transition that would be seen.

I do not propose to demonstrate the existence of this transition here. Rather, I will focus on what I have assumed is a necessary condition for its existence: that both my formulas and Koop's are the *solutions of the same equations* at different values of the same parameter $m$. To check this assumption, I have performed a parallel derivation of our formulas by using only my Trento method and equations. This means using the same Lorentz equations and the same (quadratic) approximations for both the $m = 0$ and $m > 0$ cases. (Note that the quadratic approximation for fields means a cubic approximation for potentials, if potentials are used. I use only fields.)

As I expected, using the Trento equations for $x/R$, $L/\beta$ and $d\gamma/dt$ gives Koop's results at $m=0$ and my Trento results at $m>0$. The calculations, below, are simple.

### $x/R$

My Trento equation for $x/R$ in the absence of free radial and longitudinal oscillations (TT, eq. 14) is:

$$\frac{d^2}{d\tau^2}\left(\frac{x}{R}\right) = \frac{(c/R)^2}{(1+\Delta x/R)} \times \\ \left\{(1-\vartheta_y^2)[\gamma_0^2 - 1 + 2\gamma_0 \Delta\gamma] - (\beta\gamma)_0^2[1 - m(y/R)^2][1 + \Delta\gamma/\gamma_0]\right\} \quad (2)$$

In the absence of $x$-oscillations, the left side of this eq. equals zero. By assuming $m=0$ and hence ignoring the term $(y/R)^2$, we immediately get Koop's result for $\Delta\gamma/\gamma$:

$$(\Delta\gamma/\gamma_0)[2\gamma_0^2 - (\beta\gamma)_0^2] = \vartheta_y^2(\gamma_0^2 - 1); \quad \frac{\Delta\gamma}{\gamma_0} = \vartheta_y^2 \frac{\gamma_0^2 - 1}{\gamma_0^2 + 1}. \quad (K1)$$

Assuming $0 < m \ll 1$ and averaging eq. (2) over times much bigger than the period of the vertical oscillations, we almost immediately also get my Trento result for $\Delta\gamma/\gamma$: $\overline{(\gamma - \gamma_0)} = 0$ (TT, page 16; TN, eq. 14'). We cannot ignore the fact that our radial electric field (TT, eq. 5) depends quadratically on a particle's vertical position, see term $m(y/R)^2$ in (2) and TT, eq. 14. But we can ignore the nonlinear term $E_0 mxy/R^2$ of our vertical electric field (see TT, eq. 5) because the free $x$-oscillations are absent by assumption and the product $y\Delta x$ is cubic. The equation for vertical deviation $y$ —not for $x$— in the quadratic approximation then becomes linear. For linear oscillations,

$$\langle \vartheta_y^2 \rangle = \langle \dot{y}^2/c^2\beta^2 \rangle = m\langle (y/R)^2 \rangle. \quad (3)$$

From eqs. (3) and (TT, eq. 14):

$$0 = \frac{(c/R)^2}{(1+\Delta x/R)}\{2\gamma_0 \Delta\gamma - (\beta\gamma)_0^2 \Delta\gamma/\gamma_0\}, \qquad \langle \Delta\gamma/\gamma_0 \rangle = 0. \tag{O1}$$

## $L/\beta$

Now we turn to the Trento equation for synchrotron stability (TT, eq. 12):

$$\langle L/\beta \rangle = L_0/\beta_0, \tag{4}$$

where $L$ is the length of the actual trajectory and $\beta$ is the velocity along this actual trajectory, with $\Delta\beta/\beta = (\Delta\gamma/\gamma)/(\gamma^2 - 1)$. This gives Koop's formula for $\Delta x/R$ in the $m = 0$ case:

$$\frac{(1+\Delta x/R)(1+\vartheta_y^2/2)}{[1+\vartheta_y^2/(\gamma_0^2+1)]} = 1, \qquad \Delta x/R = -\frac{\vartheta_y^2}{2}\frac{\gamma_0^2-1}{\gamma_0^2+1}. \tag{K2}$$

And for my $m > 0$ case, $1/N^2 \ll m \ll 1$ (TT, page 19; TN, eq. 12'):

$$(1+\Delta x/R)(1+\vartheta_y^2/2) = 1, \qquad \langle \Delta x/R \rangle = -\langle \vartheta_y^2/2 \rangle = -m\langle y^2/2R^2 \rangle. \tag{O2}$$

## $d\gamma/dt$

Finally, we must check the compliance of the (K1), K2) and (O1), (O2) with the Trento formula for $\Delta\gamma/\gamma$ (TT, eq. 13), which is

$$\frac{\Delta\gamma}{\gamma_0} = \frac{\delta\gamma}{\gamma_0} - \beta_0^2 \frac{\Delta x}{R} - \beta_0^2 m \frac{y^2}{2R^2}. \tag{5}$$

Here, $\Delta\gamma/\gamma_0$ is either (K1) at $m = 0$, or (O1) at $m > 0$; $\Delta x/R$ is either (K2) at $m = 0$ or (O2) at $m > 0$.

So far, so good. But what is $\delta\gamma/\gamma_0$ in eq. (5)? As explained in [2], this equation is the integral of (TT, eq. 2):

$$d\gamma/d\tau = (e/mc)\left[E_R dx/d\tau + E_V dy/d\tau + E_L ds/d\tau\right], \qquad (6)$$

which describes the work of electric forces. The "non-electric" term $\delta\gamma/\gamma_0$ in eq. (5) is *added after* the integration and equals a particle's kinetic energy existing independently of electric forces. In the $m=0$ case, it is the kinetic energy of the particle's vertical movement with a constant initial velocity, $\dot{y}_0 = \dot{s}_0 \cdot \vartheta_{y0}$.

Thus,

$$\delta\gamma/\gamma_0 = \frac{\gamma_0^2 - 1}{2\gamma_0^2}\vartheta_y^2, \quad m=0. \qquad (7)$$

From (K 2) and (7),

$$\Delta\gamma/\gamma_0 = \delta\gamma/\gamma_0 - \beta_0^2 \Delta x/R = \frac{(\gamma_0^2-1)}{2}\vartheta_y^2\left[\frac{1}{\gamma_0^2} + \frac{\beta_0^2}{\gamma_0^2+1}\right] = \frac{\gamma_0^2-1}{\gamma_0^2+1}\vartheta_y^2, \qquad (8)$$

which is (K1). Koop's formulas for the $m=0$ case therefore satisfy all the relevant Trento equations.

Since it is obvious that $\delta\gamma/\gamma_0 = 0$ in the $m>0$ case, my formulas satisfy eq. (5). Therefore, they also satisfy all the relevant Trento equations. (In the absence of free radial and longitudinal oscillations, the remaining unconsidered Lorentz equation—for longitudinal motion—does not change this conclusion.)